# Magnetic nanoparticles as efficient bulk pinning centers in type-II superconductors

Alexey Snezhko[1‡], Tanya Prozorov[1,2] and Ruslan Prozorov[1*]

[1] *Department of Physics & Astronomy and NanoCenter, University of South Carolina, Columbia, SC 29208*

[2] *School of Chemical Sciences, University of Illinois at Urbana-Champaign, Urbana, IL 61801*

## ABSTRACT

Enhancement of vortex pinning by magnetic nanoparticles embedded into the bulk of type–II superconductor is studied both theoretically and experimentally. Magnetic part of the pinning force associated with the interaction between a finite-size spherical magnetic inclusion and an Abrikosov vortex was calculated in London approximation. Calculations are supported by the experimental results obtained on sonochemically modified $MgB_2$ superconductor with embedded magnetic $Fe_2O_3$ nanoparticles and compared to $MgB_2$ with nonmagnetic $Mo_2O_5$ pinning centers of similar concentration and particle size distribution. It is shown that ferromagnetic nanoparticles result in a considerable enhancement of vortex pinning in large-κ type-II superconductors.

---

[‡] Presently at *Materials Science Division, Argonne National Laboratory, Argonne, IL 60439*

[*] Corresponding author: e-mail: prozorov@sc.edu. fax: 803-777-3065. tel: 803-777-8197

## Introduction

Practical applications of a superconductor are determined by the amount of electric current it can carry without energy dissipation. However, when a magnetic field on the surface of a type-II superconductor exceeds first critical filed, $H_{c1} = \left(\ln(\kappa) + 0.5\right)\Phi_0 /(4\pi\xi^2\kappa^2)$, formation of Abrikosov vortices becomes energetically favorable [1-3]. Here $\Phi_0 \approx 2 \times 10^{-7}\,\text{G}\cdot\text{cm}^2$ is the flux quantum, $\kappa = \lambda/\xi$ is the Ginsburg-Landau parameter, $\lambda$ and $\xi$ are London penetration depth and coherence length, respectively. For high-$T_c$ superconductors, typical values of $\kappa$ are large, $\kappa > 100$, hence corresponding first critical field is small. In the presence of a supercurrent density $\vec{j}$, vortices experience a Lorentz force, $\vec{F}_L = \left[\vec{j} \times \vec{\Phi}_0\right]/c$. If nothing hinders their motion, vortices move and reach some viscosity-limited velocity $\vec{v}$. This motion creates an electric field, $\vec{E} = \left[\vec{B} \times \vec{v}\right]/c$, parallel to the current. This results in a finite resistivity, $\rho = E/j$, hence energy dissipation. In order to prevent such process, vortices should be immobilized (pinned) [1-3]. Pinning of vortices on structural inhomogeneities is a common way to increase critical current density. In a uniform type-II superconductor energy of a normal phase is higher by a factor of $H_c^2/(8\pi)$ per unit volume. Therefore, a vortex with the normal core of size $\xi$ has additional energy $H_c^2\xi^2/8$ per unit length. If such vortex occupies a non-superconducting defect, part of the condensation energy is recovered and therefore this defect represents a potential well with respect to the vortex displacement. Due to its origin, this is known as condensation energy or core pinning. There are various ways to introduce bulk pinning centers of different nature, concentration, distribution and geometry to better utilize the condensation energy pinning (see, e.g., [1-3] for review).

However, there is another possibility for the enhancement of pinning - direct magnetic interaction of vortices with *ferromagnetic* pinning centers. This idea was explored already in 60s when pinning enhancement was clearly demonstrated in low-temperature superconducting alloys with magnetic nanoparticles mechanically mixed in [4-6]. More recent works concentrated on ferromagnetic particles deposited on the surface [7-10] or in the surface layer [11] of low-temperature superconductors. Martin *et al.* have successfully used a lattice of ferromagnetic dots (Fe or Ni) to create a periodic array of artificial magnetic pinning centers on superconducting Nb films [8]. Moschalkov *et al.* and Van Bael *et al.* studied submicron Co particles placed on a thin Pb film and their influence on the magnetic response of a superconductor [9,10]. It was found that periodic lattice of magnetic particles acts as an efficient 2D pinning array with pronounced matching effects. Motivated by these experiments, theoretical models of magnetic and transport responses of superconducting films with magnetic particles placed on the surface were developed [12,13]. Although being conceptually important, these studies focused on a specific case of magnetic particle on (or close to) the surface of thin superconducting film. The question of bulk magnetic pinning remained unexplored. In a related study, Rizzo *et al.* used ferromagnetic nanoparticles embedded in NbTi wires to achieve larger pinning strength compared to nonmagnetic Ti impurities [21]. The enhancement was attributed to the absence of the proximity effect in magnetic metals and therefore more efficient suppression of the superconducting order parameter. However, this mechanism does not work in case of non-metallic pinning centers (such as oxides studied in this work) and would only be efficient for nanoparticles smaller than the coherence length (about 5 nm for $MgB_2$).

In this work, *bulk magnetic pinning* due to direct magnetic interaction is studied both theoretically and experimentally. A (ferro- or ferri-) magnetic nanosized inclusion in the bulk of a type-II superconductor acts not only as a conventional core - pinning center, but also gives rise to an additional magnetic component of pinning via direct magnetic interaction with vortices. Surprisingly, there is no extended theory available yet for the description of such system. Our model consists of an infinite type-II superconductor containing an isolated straight Abrikosov vortex and a spherical magnetic particle. Pinning force associated with the magnetic interaction between a vortex and a particle is calculated in London approximation. The calculations are supported by the experimental results obtained on sonochemically modified $MgB_2$ nanocomposites with inclusions of nanosized magnetic particles of $Fe_2O_3$ and compared to nonmagnetic $Mo_2O_5$ pinning centers with similar concentration and size distribution of nanoparticles.

Introduction of ferromagnetic nanoparticles into the bulk of ceramics without agglomeration and/or significant phase separation is a non-trivial task. Recently, sonochemical method for modification of granular superconductors and *in-situ* production of magnetic pinning centers has been developed [16]. In liquid-powder slurries irradiated with high-intensity ultrasound, acoustic cavitation induces turbulent flow and shock waves. The implosive collapse of bubbles during cavitation results in extremely high local temperatures and pressures, and stimulates high-velocity collisions between suspended particles. The estimated speed of colliding particles approaches half of the speed of sound in the liquid. Effective temperatures at the point of impact can reach 3000 K, and ultrasound-caused interparticle collisions are capable of producing localized inter-particle melting and "neck" formation [14,15]. Irradiation of powdered slurries in the presence of volatile organometallics precursors produces material with nanoparticles embedded in the bulk of irradiated powders [16]. In particular, sonication of $MgB_2$ slurry in decane with addition of small amount of $Fe(CO)_5$ yields $MgB_2$-$Fe_2O_3$ nanocomposite with significantly enhanced vortex pinning [16]. Apparently, ultrasonic irradiation in the powder slurries in the presence of volatile organometallics combines the effects from both extreme cavitational hot-spot and the shock-waves generated in the liquid upon implosive bubble collapse.

## Magnetic inclusion in the bulk of type-II superconductor

Let us consider a spherical magnetic particle of radius $R$ and magnetization $\vec{M}$, embedded into an infinite type-II superconductor containing a single straight vortex line at the distance $d$ from the center of the particle. This geometry is illustrated in Fig.1, where $\vec{\Phi}_0$ indicates the direction of a magnetic field in the vortex, which carries a flux quantum $\Phi_0$. $\alpha$ is the angle between $\vec{M}$ and $\vec{\Phi}_0$. To calculate the pinning force associated with vortex - nanoparticle interaction, distributions of magnetic induction and screening currents induced by the magnetic particle must be calculated. We use the London equation for the vector-potential $\vec{A}$ ( $\nabla \times \vec{A} = \vec{B}$, $\nabla \vec{A} = 0$ ) in a superconductor:

$$\vec{A} - \lambda^2 \Delta \vec{A} = 0$$

and the Maxwell equations inside the magnetic particle:

$$\Delta \vec{A} = 0$$

Due to symmetry of the problem it is more convenient to solve the equations in spherical coordinate system $(\rho, \varphi, \theta)$ with the direction $\theta = 0$ parallel to the magnetization vector $\vec{M}$ of a particle. In this case vector-potential $\vec{A}$ has only one component $(0, A_\varphi(\rho, \theta), 0)$ and corresponding equations become:

$$\begin{cases} \dfrac{\partial^2 A_\varphi}{\partial \rho^2} + \dfrac{2}{\rho}\dfrac{\partial A_\varphi}{\partial \rho} + \dfrac{\cos\theta}{\rho^2 \sin\theta}\dfrac{\partial A_\varphi}{\partial \theta} + \dfrac{1}{\rho^2}\dfrac{\partial^2 A_\varphi}{\partial \theta^2} - A_\varphi \left( \dfrac{1}{\rho^2 \sin^2\theta} + \dfrac{1}{\lambda^2} \right) = 0, & \rho \geq R \\ \dfrac{\partial^2 A_\varphi}{\partial \rho^2} + \dfrac{2}{\rho}\dfrac{\partial A_\varphi}{\partial \rho} + \dfrac{\cos\theta}{\rho^2 \sin\theta}\dfrac{\partial A_\varphi}{\partial \theta} + \dfrac{1}{\rho^2}\dfrac{\partial^2 A_\varphi}{\partial \theta^2} - \dfrac{A_\varphi}{\rho^2 \sin^2\theta} = 0, & \rho < R \end{cases}$$

The solution should satisfy the following boundary conditions: vector potential and tangential components of the magnetic field must be continuous on the particle's surface,

$$A_\varphi^{sc}\big|_{r=R} = A_\varphi^{m}\big|_{r=R}$$

$$H_t^{sc}\big|_{r=R} = H_t^{m}\big|_{r=R} \Rightarrow \left(\nabla \times \vec{A}^{sc}\right)_\theta\big|_{r=R} = \left(\nabla \times \vec{A}^{m} - 4\pi \vec{M}\right)_\theta\big|_{r=R}$$

Here *m* stands for the solution inside magnetic sphere and *sc* denotes superconductor. In addition, vector potential should vanish inside a superconductor at $\rho \to \infty$ and be finite inside the magnetic sphere. Then, the above equations have the following solutions:

$$\begin{cases} A_\varphi = \dfrac{4\pi M R \sin\theta}{\left(1 + 3(\lambda/R) + 3(\lambda/R)^2\right)} \dfrac{(1 + \rho/\lambda)}{(\rho/\lambda)^2} \exp\left(-\dfrac{(\rho - R)}{\lambda}\right), & \rho \geq R \\ A_\varphi = \dfrac{4\pi M \rho \sin\theta}{\left(1 + 3(\lambda/R) + 3(\lambda/R)^2\right)} \dfrac{(1 + R/\lambda)}{(R/\lambda)^2}, & \rho < R \end{cases}$$

Corresponding screening current induced by the magnetic sphere has only one component $(0, j_\varphi(\rho, \theta), 0)$ and is calculated from the vector potential via $4\pi c^{-1} j_\varphi = (\nabla \times \vec{\mathbf{B}})_\varphi = -(\Delta \vec{\mathbf{A}})_\varphi = -A_\varphi \lambda^{-2}$. Therefore, supercurrent induced around the magnetic sphere is given by:

$$j_\varphi = -\frac{cMR}{\left(1+3(\lambda/R)+3(\lambda/R)^2\right)} \frac{(1+\rho/\lambda)}{\rho^2} \exp\left(-\frac{\rho-R}{\lambda}\right) \sin\theta$$

Let us now calculate the pinning force associated with the magnetic interaction between a magnetic sphere and a vortex placed nearby. Assuming that vortex is positioned at a distance $d$ from the particle, the corresponding interaction force is calculated as:

$$\vec{\mathbf{F}}_{mag} = c^{-1} \int \left[\vec{\mathbf{j}}(\rho_v,\theta_v) \times \vec{\mathbf{\Phi}}_0\right] d\mathbf{l}$$

where $d\mathbf{l}$ is a flux line element, $c$ is the speed of light, $\vec{\mathbf{j}}(\rho_v,\theta_v)$ is the supercurrent density at the location of the vortex core, and the integration is carried over the entire vortex length. Evidently, the resulting force is attractive and maximal when particle's magnetization and magnetic field of a vortex are collinear. In a general case of an arbitrary orientation, the value of magnetic pinning force is scaled by the factor of $\cos(\alpha)$, where $\alpha$ is the angle of misalignment. In addition, the vortex is experiencing additional moment of forces which is trying to align it along the magnetic moment of the sphere. The magnitude of this moment of forces, $K$, acting on the vortex line with respect to the point where vortex crosses the θ=π/2 plane is given by:

$$K = 2M\Phi_0 R \sin(\alpha) \frac{\exp\left(\frac{R}{\lambda}\right)}{1+3(\lambda/R)+3(\lambda/R)^2} \mathrm{P}\left(\frac{d}{\lambda}\right)$$

$$\mathrm{P}\left(\frac{d}{\lambda}\right) = \int_{d/\lambda}^{\infty} (1+x) x^{-2} \sqrt{x^2 - \left(\frac{d}{\lambda}\right)^2} \exp(-x)\, dx,$$

Calculated magnetic pinning force acting on the flux line placed near the magnetic spherical particle, is shown in Figure 2 for different particle radii. The Ginsburg-Landau parameter, $k = \lambda/\xi$, was chosen to be 100 in our calculations. The evolution of the moment of forces acting on the vortex due to misalignment with magnetization vector of particle is presented in the inset as a function of the distance between vortex and particle. For angles $\pi/2 < \alpha < \pi$, magnetic pinning force becomes repulsive. However, for a large number of nanoparticles randomly distributed in the bulk of a superconductor, even repulsive forces leads to an enhancement of the bulk pinning force [1-3].

## Experimental

Detailed description of ultrasonic modification of $MgB_2$ and sonochemical preparation of $MgB_2$-$Fe_2O_3$ superconductor-ferromagnet nanocomposites has been reported elsewhere [16]. In brief, 2% wt slurry of $MgB_2$ polycrystalline powder (325 mesh, Alfa Aesar) in 15 ml of decalin was irradiated with ultrasound at 20 kHz and ~50 W/cm$^2$ under ambient atmosphere, using direct-immersion ultrasonic horn (Sonics VCX-750). $MgB_2$-$Fe_2O_3$ nanocomposites were prepared by sonochemical irradiation of 2 %wt $MgB_2$ slurry with the addition of 0.5 mmol $Fe(CO)_5$, 0.9 mmol $Fe(CO)_5$, and 1.8 mmol $Fe(CO)_5$, respectively [17]. $MgB_2$-$Mo_2O_5$ nanocomposites were prepared by sonochemical irradiation of 2% wt $MgB_2$ slurry with the addition of 0.5 mmol $Mo(CO)_6$, 0.9 mmol $Mo(CO)_6$ and 1.8 mmol $Mo(CO)_6$, respectively [18,19]. The resulting materials were filtered, washed repeatedly with pentane, and air-dried overnight. No post-synthetic annealing was performed, since we intended to rule out net effect of ultrasonic irradiation on the material. Equal amounts of $Fe(CO)_5$ or $Mo(CO)_6$ were added in each case, therefore sonochemical synthesis yielded similar number and size distribution of in-situ produced nanoparticles [20]. Scanning electron microscopy study was conducted on Hitachi S-4700 instrument. Average size of agglomerates obtained after irradiation of granular $MgB_2$ with high-intensity ultrasound, was found to be ~ 30 μm. Samples were additionally characterized by powder X-ray diffraction. Figure 3 shows SEM images of (A) polycrystalline $MgB_2$ before sonication, (B) $MgB_2$ irradiated with high-intensity ultrasound, (C) $MgB_2$ sonicated with $Fe(CO)_5$, which is believed wo produce a nanocomposite material with $Fe_2O_3$ nanoparticles [17,20], and (D) $MgB_2$ sonicated with $Mo(CO)_6$ – resulted, presumably, in embedded $Mo_2O_5$ nanoparticles [18,19]. No nanoparticles were formed in $MgB_2$ sonicated without any organometallic compounds. In contrast, sonication of granular $MgB_2$ in the presence of volatile organometallic compounds allows formation of nanoparticles (brighter ~50 nm spots easily visible in last two pictures). In order to verify that these are, indeed, nanoparticles obtained *in-situ*, the localized energy dispersive X-ray analysis (EDX) was conducted on these materials, using both scan mode and spot-mode. Figure 4 (A) shows EDX spectra obtained in nanocomposite containing $Fe_2O_3$ measured off the spot as indicated on the corresponding SEM image on the right. The relative content of iron with respect to magnesium is nearly zero. The situation is opposite for the on-spot measurement as shown in Figure 4 (B). There, iron oxide with the nominal chemical composition of $Fe_{1.8}O_{3.1}$ is detected. Similar results are obtained for the nanocomposites containing $Mo_2O_5$ nanoparticles. Figure 4 (C) nanocomposite off-spot measurement and Figure 4 (D) shows on-spot measurement for the material, indicating formation of oxide species with the nominal chemical composition of $Mo_{1.9}O_{4.9}$. Composition of metal oxide nanoparticles determined with EDX measurements essentially matches the stoichiometric composition of iron and molybdenum oxides, $Fe_2O_3$ and $Mo_2O_5$. Traces of titanium (Figure 4 (A, C)) are attributed to the erosion of Ti-horn caused by the abrasive action of suspended of $MgB_2$ grains during the ultrasonic irradiation of decalin slurries.

Magnetic measurements were conducted using *Quantum Design* Superconducting Quantum Interference Device (SQUID) MPMS magnetometer. The average sample mass was 10 mg. Measured magnetic moment was normalized using the initial slope, $dM/dH$. This slope is proportional to the volume of the superconducting phase, and for materials without magnetic nanoparticles such normalization eliminates the contribution of the demagnetization factor and gives the volume magnetization. For composites containing $Fe_2O_3$ nanoparticles, the normalization was done after subtraction of the paramagnetic contribution measured above $T_c$, which however was almost negligible. In Figure 5, temperature dependence of magnetization measured in a magnetic field of 1 kOe applied after zero-field cooling is shown for all three samples studied: $MgB_2$ (sonicated, no additives); $MgB_2$ containing $Fe_2O_3$ nanoparticles (obtained by sonication with $Fe(CO)_5$ and $MgB_2$ with nonmagnetic $Mo_2O_5$ nanoparticles (obtained by sonication with $Mo(CO)_6$. Figure 6 shows magnetization loops measured at 5 K in the same samples. As expected, magnetization loops for $MgB_2$ with nanosized inclusions are more hysteretic compared to the material without inclusions, which implies enhanced pinning strength. It should be noted that superconducting transition temperature did not change after sonochemical treatment, see Fig. 5. This indicates that magnetic component does not act as a dopant, but forms well shaped inclusions, also observed by SEM and TEM and confirmed by EDX, Figure 4. Similar magnetic behavior was observed for other concentrations of embedded nanoparticles. Comparing the samples with nonmagnetic and magnetic pinning centers, we conclude that magnetic nanoparticles lead to a considerable increase of the total bulk pinning force. Another important conclusion is that magnetic pinning is more efficient in high-$\kappa$ high-$T_c$ superconductors compared to low-$\kappa$ low-$T_c$ superconductors. This is because the energy of a vortex in type-II superconductor can be written as: $E_v \approx \left(\Phi_0/4\pi\lambda\right)^2 \left(\ln\kappa + 0.5\right)$, where additional factor 0.5 to $\ln\kappa$ comes from the contribution of the vortex core [3]. For low-$\kappa$ superconductors, core-energy term is dominant or comparable to the magnetic term. Consequently, lowering the magnetic energy by interaction with a magnetic inclusion does not significantly affect total vortex energy. As a result, the effectiveness of magnetic pinning is relatively low. The situation is opposite in high-κ materials, such as high-temperature superconductors where magnetic term is dominant and minimization of the magnetic vortex energy significantly lowers its total energy resulting in higher effectiveness of pinning associated with the magnetic interaction. Another interesting aspect of magnetic pinning is its dependence on the angle between the direction of magnetization in nanoparticles and the flux lines. Oriented nanocomposite materials should have anisotropic pinning enhancement. The experimental work on oriented nanocomposites is in progress.

## Conclusions

In conclusion, it is found that magnetic pinning force is a long-range attractive force with a characteristic length λ. The magnitude of this force depends on the magnetization value, particle size and orientation of the magnetization vector in the magnetic particle with respect to the flux line. The experiments with $MgB_2$ superconductor treated with high-intensity ultrasound have confirmed the theory. Considerable improvement of a magnetic hysteresis was observed for samples with embedded magnetic $Fe_2O_3$ nanoparticles compared to non-magnetic $Mo_2O_5$ of similar concentration and size distribution. Our results suggest a new direction in the improvement of vortex pinning in high-$T_c$ superconductors.

## **Acknowledgements**

Discussions with Yu. Genenko, B. Ivlev, V. Geshkenbein, and K. S. Suslick are greatly appreciated. This work is supported by the NSF-EPSCoR Grant EPS-0296165, a grant from the University of South Carolina Research and Productive Scholarship Fund, and the donors of the American Chemical Society Petroleum Research Fund. The SEM study was carried out in the Center for Microanalysis of Materials (UIUC), which is partially supported by the DOE under Grant DEFGO2-91-ER45439.

**REFFERENCES**

FIGURES:

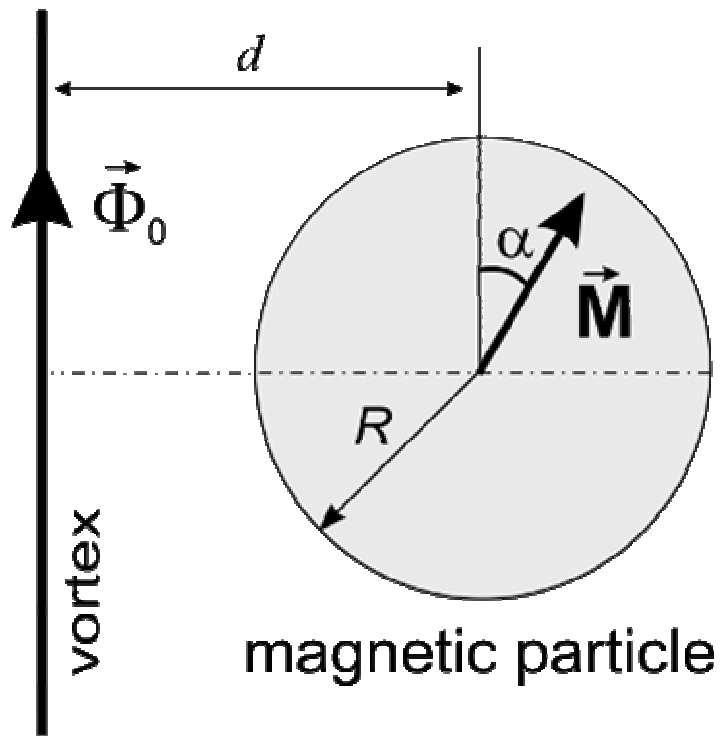

Figure 1. Geometry of calculations: an isolated vortex at a distance *d* from a magnetic sphere of radius *R* and magnetization $\vec{M}$. $\vec{\Phi}_0$ indicates the direction of a magnetic field in the vortex carrying flux quantum $\Phi_0$.

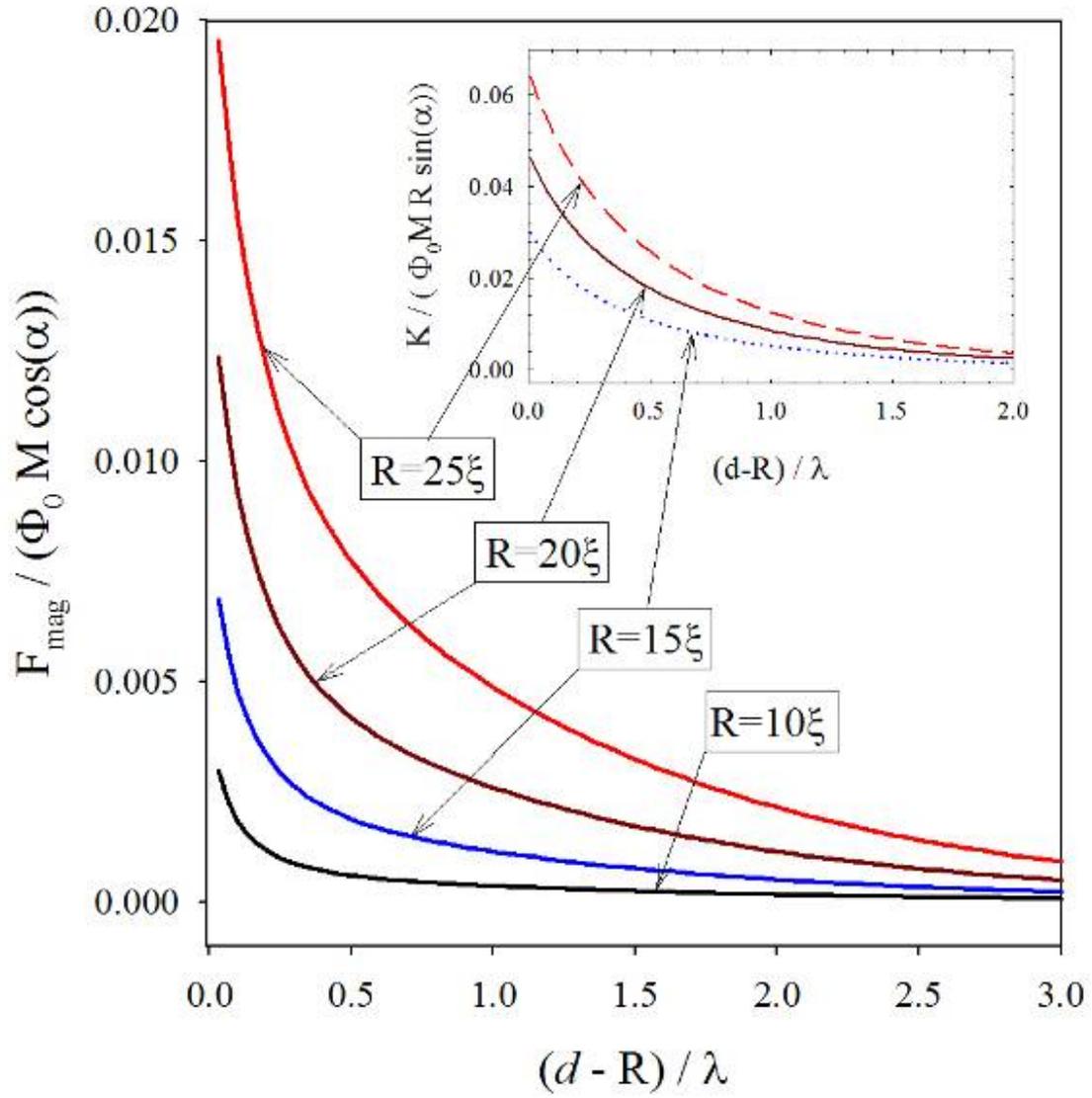

Figure 2. Magnetic pinning force as a function of the distance between a vortex and a magnetic sphere calculated for different sphere radiuses for $\kappa = 100$. *Inset:* Variation of the moment of forces, *K*, with the vortex-particle distance for different particle's radiuses.

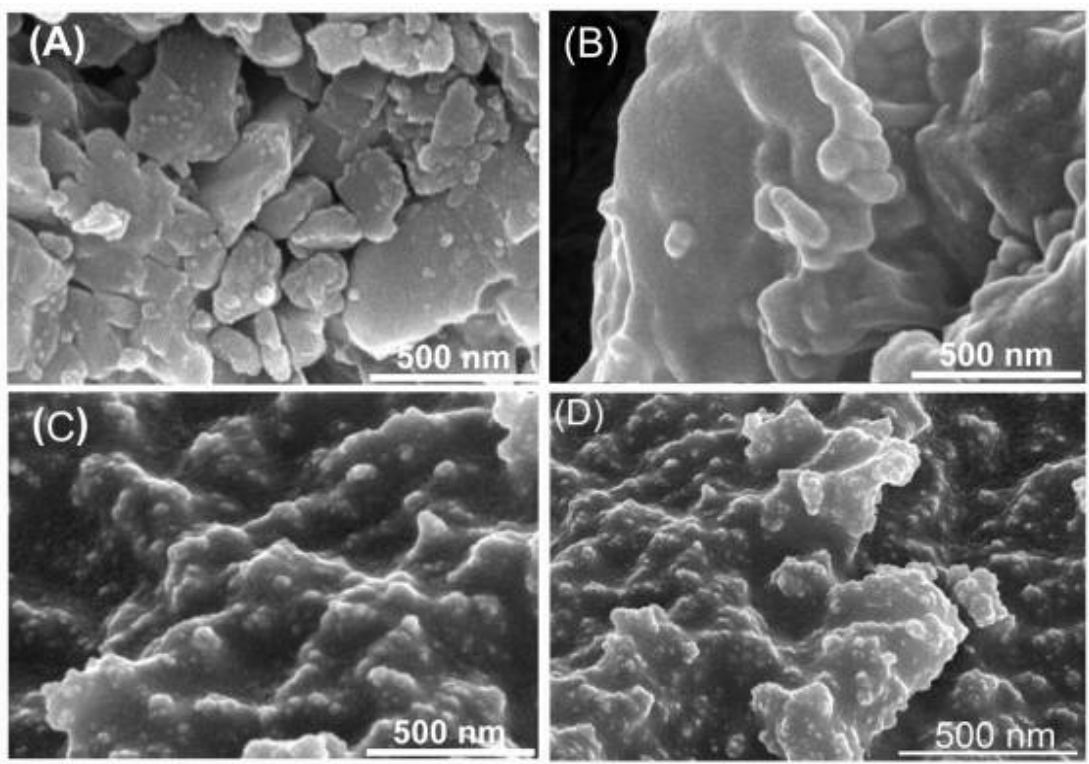

Figure 3. Sonochemical modification of $MgB_2$ superconductor: **(A)** Starting material. **(B)** $MgB_2$ irradiated with high-intensity ultrasound. **(C)** $MgB_2$ sonicated with 1.8 mmol $Fe(CO)_5$. **(D)** $MgB_2$ sonicated with 1.8 mmol $Mo(CO)_6$; Sonication was performed in 2% (w/w) decalin slurry at 263 K, 20 kHz, ~50 W/cm$^2$

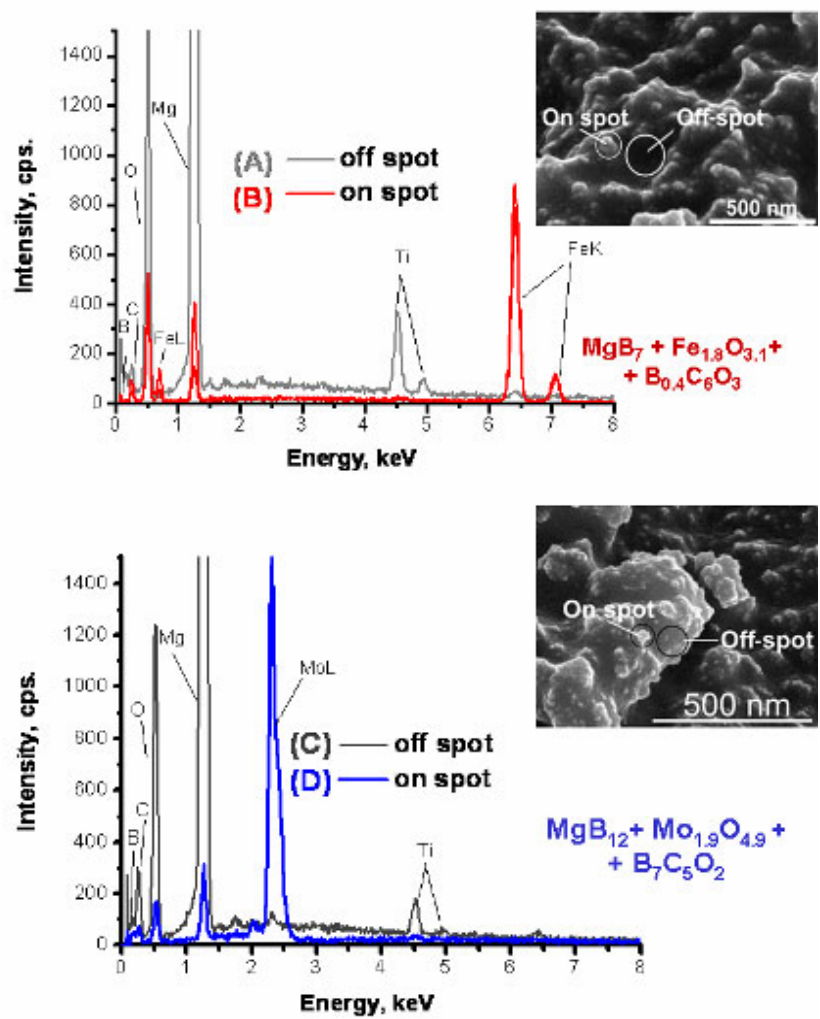

Figure 4. Local energy dispersive X-ray spectroscopy (EDX) measured in sonicated material with embedded nanoparticles. **(A)** nanocomposite containing $Fe_2O_3$ – measured off spot (see text); **(B)** same as (A), but for on-spot measurement; **(C)** nanocomposite containing $Mo_2O_5$ nanoparticles – off spot measurement **(D)** same as (C), but for on-spot measurement.

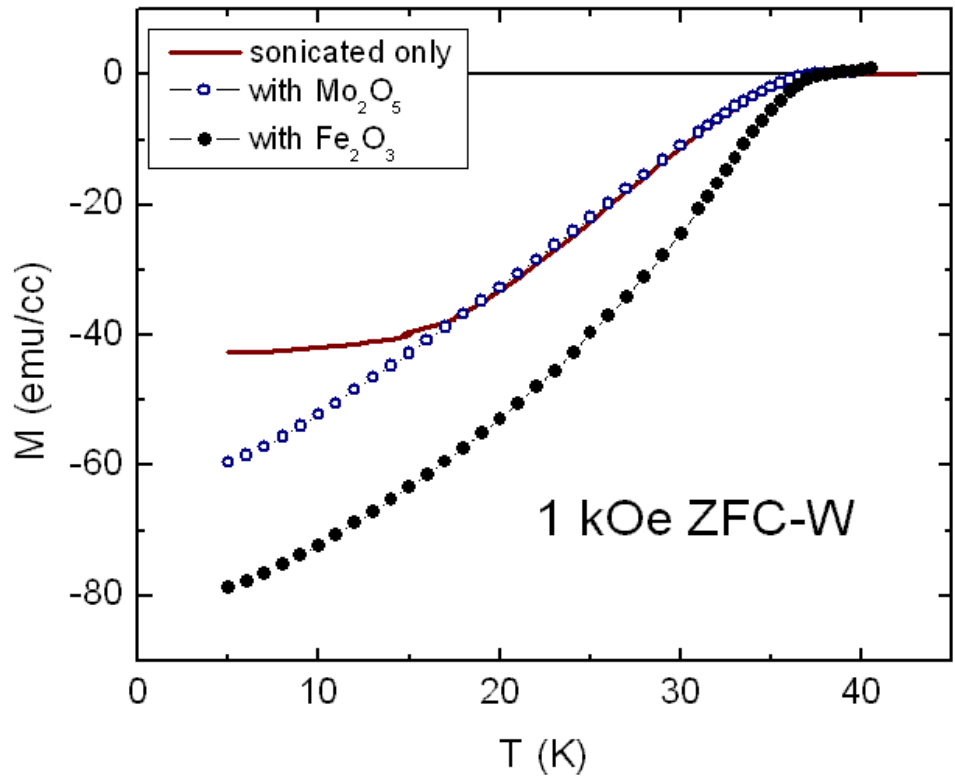

Figure 5. Temperature dependence of the magnetization measure upon application of a 1 kOe magnetic field and warming up. Solid line – reference sample; open symbols show $Mo_2O_5$ - containing nanocomposite; full symbols show nanocomposite with $Fe_2O_3$ nanoparticles.

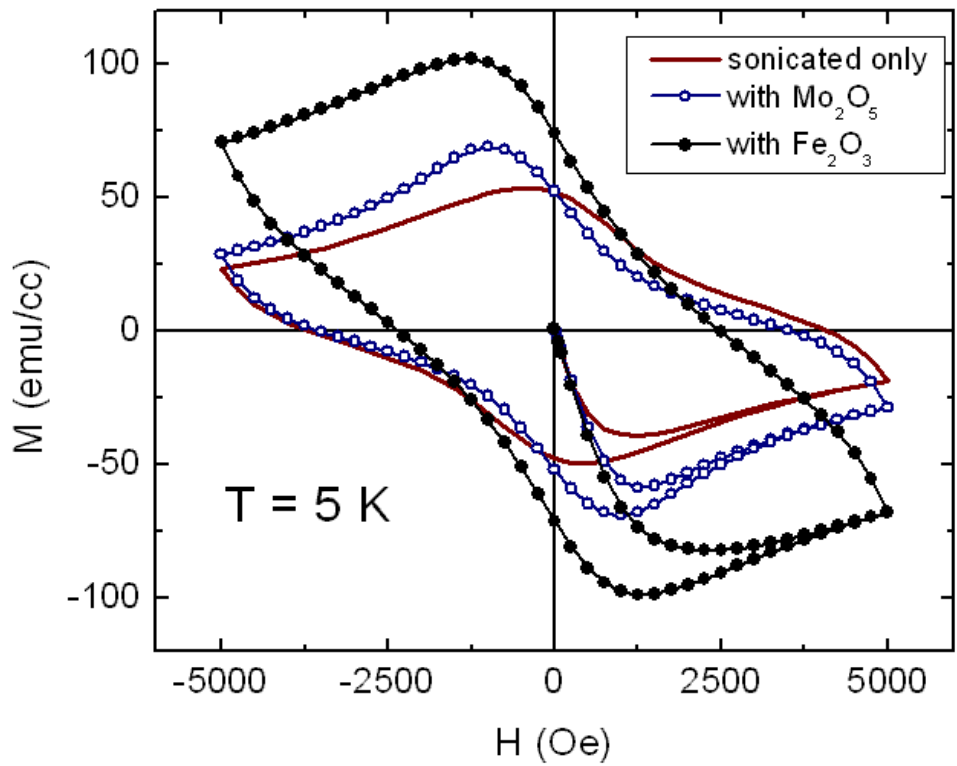

Figure 6. Magnetization loops for the reference sample (solid line), $Mo_2O_5$ - containing composites (open symbols), and $Fe_2O_3$ – containing composites (solid symbols).